# Light Weight Residual Dense Attention Net for Spectral Reconstruction from RGB Images

[a]D.Sabari Nathan, [b]K.Uma, [c]D Synthiya Vinothini, [d]B. Sathya Bama and [e]S. M. Md Mansoor Roomi
[a]AI Engineer Couger Inc, Tokyo, Japan.
[b,c]Research Scholar, [d,e]Associate Professor
[b,c,d,e]Thiagarajar college of Engineering, Madurai, India.
[b]Email id: umak@tce.edu

## Abstract

*Hyperspectral Imaging is the acquisition of spectral and spatial information of a particular scene. Capturing such information from a specialized hyperspectral camera remains costly. Reconstructing such information from the RGB image achieves a better solution in both classification and object recognition tasks. This work proposes a novel light weight network with very less number of parameters about 233,059 parameters based on Residual dense model with attention mechanism to obtain this solution. This network uses Coordination Convolutional Block to get the spatial information. The weights from this block are shared by two independent feature extraction mechanisms, one by dense feature extraction and the other by the multiscale hierarchical feature extraction. Finally, the features from both the feature extraction mechanisms are globally fused to produce the 31 spectral bands. The network is trained with NTIRE 2020 challenge dataset and thus achieved 0.0457 MRAE metric value with less computational complexity.*

## 1. Introduction

A sight that is seen by our human eye is the collection of some spectral reflectance. In ordinary cameras, the sensors along with the filters are used to convert the spectral light into three color channels. These channels can project the imaging process by replicating the scene's perceptual quality, which results in spectral information loss[1].

Fine spectral information can be observed by Hyperspectral Imaging Systems, that are used to record the spectral signature reflected from each observable point in a scene. Spectral analysis from the hyperspectral images provides richer information about the scene than a normal RGB camera. But these systems are too costly and physically larger in size to fit in portable devices[2].

Spectral Reconstruction from the RGB images provides the solution to recover the entire spectral information from each observable point in the image. The richness of spectrum information that are captured and analyzed has several benefits in various fields [3]. Hyperspectral data are extensively used in medical field[4,5,6], astronomy[7] and remote sensing [8].

### 1.1 Related Works

Hyperspectral Imaging Devices are developed to capture the high resolution radiance spectra at every pixel in an image. Recent development in this technology looks for faster image capturing compared to the conventional scanning based techniques [9]. In the existing technologies, the hyperspectral images are directly captured which includes prism-mask system, a multispectral video acquisition method with different spatial and spectral resolution [10] and by using multiple cameras in different spectral sensitivities [11,12]. Instead of designing a new setup to capture hyperspectral images, spectral reconstruction from the RGB image provides better solution. Sparse coding provides the simple model and fast training for reconstruction[13]. Of the recent technologies, the deep neural networks achieve a leading performance in spectral reconstruction, Generative Adversarial Networks[14], 2D and 3D CNN model for the spectral data reconstruction[15,16,17] Manifold learning based method[18] and metameric spectral Reconstruction [19].

Hyperspectral images are limitedly available and are very hard to download. Collection of such hyperspectral image dataset also lacks spatial resolution. To overcome the above problem, this work provides a cost effective solution for hyperspectral imaging that can reconstruct the spectral information from the RGB image.

The main contributions of the proposed network can be summarized below:

(i) A Lightweight Model is proposed with very less number of parameters about 233,059 parameters.

(ii) The Residual Dense Block is incorporated with spatial and channel attention mechanism to extract significant local hierarchical features. Such RDAB blocks are connected at multiscale to extract multiscale hierarchical features.
(iii) Weight sharing is incorporated in the input and feature fusion at the output.
(iv) The structural similarity loss function is used for training the network, this ensures that the reconstructed image preserves the required structural features from the input.
(v) Experiments were conducted on the NTIRE 2020 spectral reconstruction clean dataset and confirmed the efficiency of the proposed architecture.

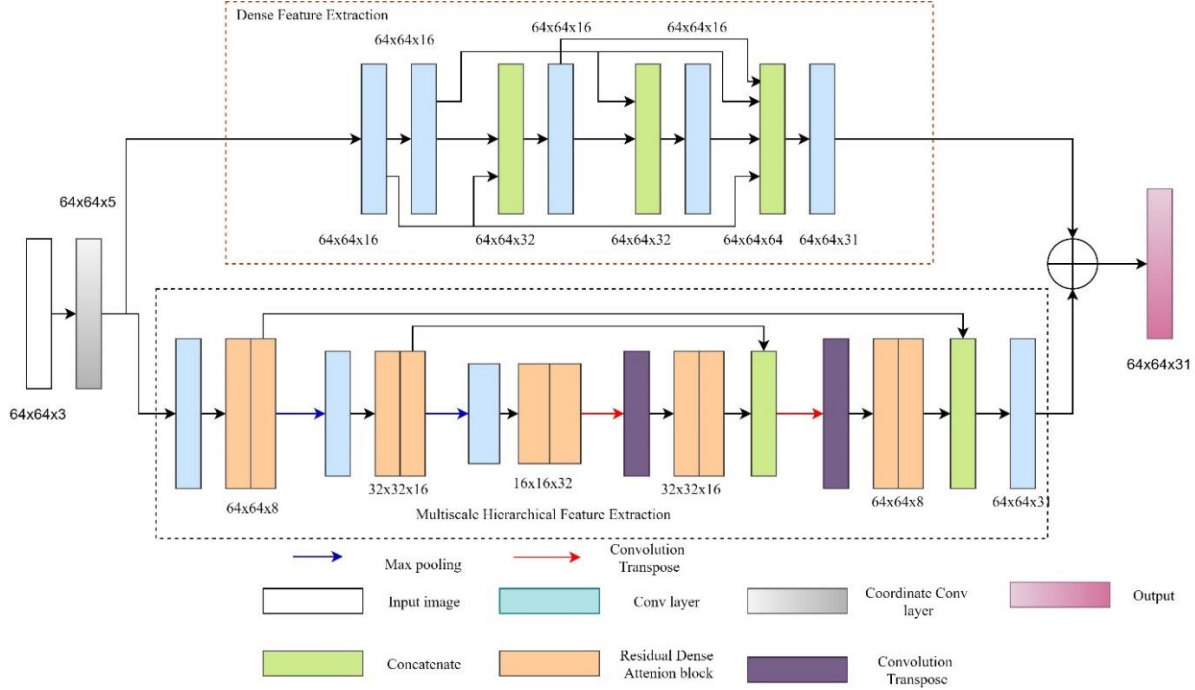

Figure 1: Spectral Reconstruction Architecture from RGB images

The rest of this paper is organized as follows: Section II deals in detail about the network architecture and the process flow involved in training and testing. Section III presents the experimental results, quality measurements and its analysis. Finally, section IV concludes with a brief summary.

## 2. Network Structure

A high spatially resolved RGB image is taken as input. To obtain its coordinate information in both parallel and perpendicular axes and to determine the pixel position in the feature maps, coordination convolutional block is used. To extract such features, the initial local features extracted from the CoordConv layer are down sampled to 16*16 and passes through the network to form a bottleneck. The proposed Architecture consists of two phases for dense feature extraction and multiscale hierarchical feature extraction. The Dense features are extracted by a dedicated dense connection of convolutional and concatenation layer of size 64*64. The multiscale hierarchical features are extracted by the Residual Dense Attention Block (RDAB). The proposed Network with an ensemble of convolution layer and Residual Dense Attention Block connected at multi-scale level, used for spectral reconstruction is shown in Figure 1. Specifically, in each RDAB block, certain significant features are given more importance spatially and spectrally by its dedicated attention mechanism; henceforth multi-scale hierarchical features are extracted at multi-level to widen the forward paths for higher capacity. The Residual Dense Block (RDB) generates the local hierarchical feature. RDAB blocks are connected at multi-scale level in a U-net fashion, where the encoding phase consists of maxpooling layer in between the RDAB blocks meanwhile the decoding phase consists of Transpose Convolution between them. This Transpose Convolution helps to reconstruct the image to the same spatial resolution as that of the input. With the help of convolution transpose, the output of the RDAB is again extended to its original feature map size of 64*64. This assures that more features are available to the final 64*64 convolution layer. Finally, both the feature extraction mechanisms are fused to form a resultant of 31 spectral bands.

### 2.1 Coordination Convolution Block

Rather than the standard convolutional layer, the CoordConv Layer follows a different approach in handling spatial information and also to make the network, translation dependence throughout the training process. As shown in Figure 2 additional channels are added in the input and the values of these channels are normalized to the range from -1 to +1. Hence, it is suitable to get high boundary information and with low internal variability[20].

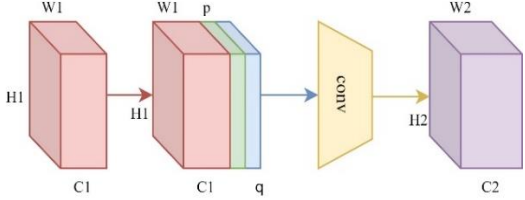

Figure 2: Coordination Convolution Block

## 2.2 Residual Dense Attention Block

The Residual Dense Attention Block (RDAB) is used to extract the local dense features associated with its channel and spatial modules is shown in Figure 3. Each block has a convolutional layer followed by ReLu activation function.

Let $F_i$ be the input and $F_o$ be the output of the RDAB block. The output of the $q$-th convolutional layer is written as

$$F_{p,q} = \beta(W_{p,q}[F_i, F_{p,1}, \ldots, F_{p, q-1}]) \quad (1)$$

where $\beta$ denotes the ReLu activation function, $W_{p,q}$ is the weight of the $q$-th convolutional layer and the bias function is excluded for simplicity. $[F_i, F_{p,1}, \ldots, F_{p, q-1}]$ is the concatenation of feature maps.

The present and previous block outputs are connected directly to the next layer to extract Local Dense Feature. Local Feature fusion is used to fuse the effective features from the previous RDB and the entire convolution layers in the current RDB block [21]. This will help the network to train in a stabilized manner. The local feature fusion is given as

$$F_{p,loc} = F_{loc}^p([F_i, F_{p,1}, \ldots F_{p,q} \ldots, F_{p,Q}]) \quad (2)$$

where $F_{loc}^p$ denotes the function LFF in the $p$-th RDB. Local Residual learning(LRL) in RDB is used to boost up the flow of information to the next block ($F_{int}$) and is given by

$$F_o = F_i + F_{p,loc} \quad (3)$$

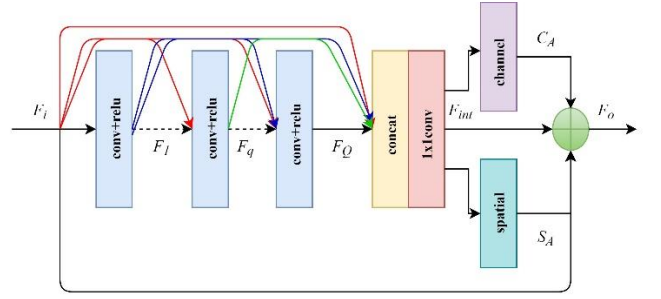

Figure 3: Residual Dense Attention Block

LRL can also further improve the network representation ability, resulting in better performance.

For refining the important features, the channel and spatial modules are incorporated in the network. Channel attention map is generated by exploiting the inter-channel relation of features whereas the spatial attention map by utilizing inter-spatial relationship of features[22].

The attention modules can be formulated as

$$ChannelAttention(C_A) = F_A\left[\omega_1\left(\omega_0\left(\frac{\sum_{k=1}^{n} x_k}{n}\right)\right) + \omega_1(\omega_0(\max(x_k)))\right] \quad (4)$$

Where $\omega_1 \in R^{c/rc}$ and $\omega_1 \in R^{c^2/r}$, $F_A$ denotes the sigmoid function.

$$SpatialAttention(S_A) = F_{conv}\left(\left[\frac{\sum_{k=1}^{n} c_{Ai}}{n} \| \max(c_{Ai})\right]\right) \quad (5)$$

$F_{conv}$ denotes the convolution operation.

The output of RDAB block is given by

$$F_o = F_i + F_{int} + C_A + S_A \quad (6)$$

## 2.3 Dense Feature Extraction Block

The dense structure for spectral reconstruction in Fig.1 can substantially alleviate the vanishing gradient problem during training. The dense connection enables the $n$-th layer to collect all the features from the previous layers and is given by

$$F_n = q_n([f_0, f_1, \ldots f_{n-1}]) \quad (7)$$

Where $q_n(.)$ denotes the $n$-th convolutional layer and $[f_0, f_1, \ldots f_{n-1}])$ is the concatenation of the feature output from the preceding layers. The concatenation operation in each dense block explicitly increases the number of channels which in turn has the ability to learn more effective model for the inverse problem.

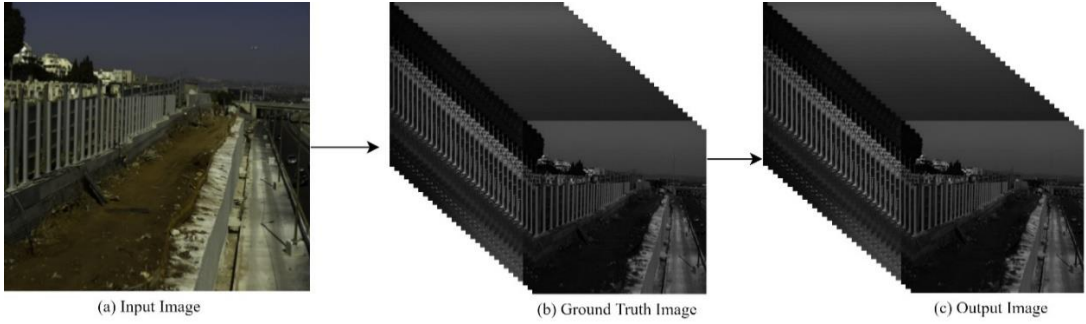

Figure 4: Experimental Results on Spectral Reconstruction

## 3. Results and Discussions

4. Table1. Performance metrics for the proposed method

### 3.1 Implementation Details

The proposed model consists of 233,059 parameters. The shared dataset consists of 460 images. We split it into 400 images for training and 50 images for validation. Training images were partitioned into sub-image patches with a resolution of 20 × 20 and a batch size of 8 was selected empirically for stochastic gradient descent. The model is trained with 8000 training patches and 1000 validation patches. Figure 4 (a) and Figure 4(b) and shows the input and ground truth images used for training. Figure 4(c) shows the resultant image reconstructed from the input RGB image. We used Adam optimizer with a learning rate of 0.001 to 0.00001 and 500 epochs for training the model. The proposed network was trained with the IntelCore i7 processor, GTX 1080 GPU, 8GB RAM, Platform keras.

### 3.2 Evaluation metrics

Structural Similarity Metric (SSIM) and Mean Relative Absolute Error (MRAE) are used for quantitative analysis of the proposed network. Let $I_R$ and $I_G$ denotes the *j-th* pixel of the reconstructed images and the ground truth hyperspectral image respectively. The evaluation metrics are calculated using the following equation

$$MRAE = \frac{1}{p}\sum_{j=1}^{p}\left(\frac{|I_R^j - I_G^j|}{I_G^j}\right) \quad (8)$$

$$RMSE = \sqrt{\frac{1}{n}\sum_{j=1}^{p}(I_R^j - I_G^j)^2} \quad (9)$$

The metrics values for the NTIRE 2020 challenge dataset is shown in Table 1. The MRAE values for the network with and without coordination convolution layer and Convolutional Block Attention Module (CBAM) are shown in Table 2 and Table 3. From these tables, it is clearly understood that using the CoordConv and the CBAM, the error rate is greatly reduced.

Table1. Performance metrics for the proposed method

| Data size | Data | MRAE | SSIM |
|---|---|---|---|
| 400 | Training Data | 0.02372 | 0.9899 |
| 50 | Validation Data | 0.04497 | 0.9827 |
| 10 | Testing Data1 | 0.05478 | - |
|  | Testing Data2 | 0.04577 | - |

Table 2. MRAE values with and without coordinate convolution

|  | With Coordinate Convolution | Without Coordinate Convolution |
|---|---|---|
| Training Data | 0.08623 | 0.2091 |
| Validation Data | 0.09933 | 0.20033 |
| Testing Data | 0.054784 | 0.09761 |

Table 3. Evaluation metrics without CBAM

|  | MRAE | RMSE |
|---|---|---|
| Training Data | 0.1432 | 0.0145 |
| Validation Data | 0.02009 | 0.121479 |
| Testing Data | 0.07594 | - |

### 3.3 Loss Function

L2 loss and SSIM loss is used to train the network. The Loss function used for the proposed Network is shown below.

$$FI_L = L_2 + L_{SSIM} \quad (10)$$

L2 loss is used here because it is very convenient property for optimization problems. Error function for a

patch is written as

$$L(P) = \frac{1}{M}\sum_{p\in P} \in (p) \qquad (11)$$

Where M is the total number of pixels in a patch.

$$L_2(P) = \frac{1}{M}\sum_{p\in P}((x(p) - y(p))^2 \qquad (12)$$

Where *p* is the pixels in the patch *P* and *x(p)*, *y(p)* are the values in the processed and the referenced images.
The loss function for SSIM can be written as

$$L_{SSIM}(P) = \frac{1}{M}\sum_{p\in P} 1 - SSIM(p) \qquad (13)$$

Substituting eqn (12) and (13) in (10) the total loss function is given by

$$FI_L = \frac{1}{M}\sum_{p\in P}[((x(p) - y(p))^2 + (1 - SSIM(p))] \qquad (14)$$

$$FI_L = \frac{1}{M}\sum_{p\in P}[((x(p) - y(p))^2 + (1 - SSIM(p))]$$

## Conclusion:

We have proposed a novel method for spectral reconstruction from a single RGB image, which gives a promising solution to low-cost and high resolution hyperspectral camera. Our approach is based on the light weight residual dense attention network which is used to extract the multi-level hierarchical features. Despite the mathematical loss of the spectral data in a RGB camera, we show that the spectral reflectance can be reconstructed with low MRAE and RMSE errors.

References

[1] R. M. H. Nguyen, D. K. Prasad, and M. S. Brown, "Training-based spectral reconstruction from a single RGB image," *Lect. Notes Comput. Sci. (including Subser. Lect. Notes Artif. Intell. Lect. Notes Bioinformatics)*, vol. 8695 LNCS, no. PART 7, pp. 186–201, 2014.

[2] B. Arad, O. Ben-Shahar, R. Timofte, L. Van Gool, L. Zhang, and M. H. Yang, "NTIRE 2018 challenge on spectral reconstruction from RGB images," *IEEE Comput. Soc. Conf. Comput. Vis. Pattern Recognit. Work.*, vol. 2018-June, pp. 1042–1051, 2018.

[3] Y. B. Can and R. Timofte, "An efficient CNN for spectral reconstruction from RGB images," 2018.

[4] D. T. Dicker *et al.*, "Differentiation of normal skin and melanoma using high resolution hyperspectral imaging," *Cancer Biol. Ther.*, vol. 5, no. 8, pp. 1033–1038, 2006.

[5] G. Lu and B. Fei, "Medical hyperspectral imaging: a review," *J. Biomed. Opt.*, vol. 19, no. 1, p. 010901, 2014.

[6] G. N. Stamatas, C. Balas, and N. Kollias, "Hyperspectral Image Acquisition and Analysis of Skin Hyperspectral Image Acquisition and Analysis of Skin," no. July, pp. 7–13, 2003.

[7] E. K. Hege, D. O'Connell, W. Johnson, S. Basty, and E. L. Dereniak, "Hyperspectral imaging for astronomy and space surveillance," *Imaging Spectrom. IX*, vol. 5159, p. 380, 2004.

[8] J. M. Bioucas-dias, A. Plaza, G. Camps-valls, P. Scheunders, N. M. Nasrabadi, and J. Chanussot, "Hyperspectral Remote Sensing Data Analysis and Future Challenges," no. june, 2013.

[9] Y.-T. Lin and G. D. Finlayson, "Physically Plausible Spectral Reconstruction from RGB Images," 2020.

[10] X. Cao, S. Member, H. Du, and X. Tong, "A Prism-Mask System for Multispectral Video Acquisition," vol. 33, no. 12, pp. 2423–2435, 2011.

[11] S. W. Oh, M. S. Brown, and M. Pollefeys, "Do It Yourself Hyperspectral Imaging with Everyday Digital Cameras," pp. 2461–2469, 2016.

[12] L. Wang, Z. Xiong, D. Gao, G. Shi, W. Zeng, and F. Wu, "High-speed Hyperspectral Video Acquisition with a Dual-camera Architecture," pp. 4942–4950, 2015.

[13] B. A. and O. Ben-Shahar, "Sparse Recovery of Hyperspectral Signal from Natural RGB Images," _ *Springer Int. Publ. AG 2016 B. Leibe al.*, vol. 1, no. 9911, pp. 19–34, 2016.

[14] A. Alvarez-Gila, J. Van De Weijer, and E. Garrote, "Adversarial Networks for Spatial Context-Aware Spectral Image Reconstruction from RGB," *Proc. - 2017 IEEE Int. Conf. Comput. Vis. Work. ICCVW 2017*, vol. 2018-Janua, pp. 480–490, 2017.

[15] S. Koundinya, H. Sharma, M. Sharma, and A. Upadhyay, "2D-3D CNN based architectures for spectral reconstruction from RGB images," *2018 IEEE/CVF Conf. Comput. Vis. Pattern Recognit. Work.*, pp. 957–9577, 2018.

[16] T. Stiebei, S. Koppers, P. Seltsam, and D. Merhof, "Reconstructing spectral images from RGB-images using a convolutional neural network," *IEEE Comput. Soc. Conf. Comput. Vis. Pattern Recognit. Work.*, vol. 2018-June, pp. 1061–1066, 2018.

[17] Z. Shi, C. Chen, Z. Xiong, D. Liu, and F. Wu, "HSCNN+: Advanced CNN-based hyperspectral recovery from RGB images," *IEEE Comput. Soc. Conf. Comput. Vis. Pattern Recognit. Work.*, vol. 2018-June, pp. 1052–1060, 2018.

[18] Y. Jia *et al.*, "From RGB to Spectrum for Natural Scenes via Manifold-Based Mapping," *Proc. IEEE Int. Conf. Comput. Vis.*, vol. 2017-Octob, pp. 4715–4723, 2017.

[19] T. Stiebel, P. Seltsam, and D. Merhof, "Enhancing Deep Spectral Super-resolution from RGB Images by Enforcing the Metameric Constraint," pp. 57–66, 2020.

[20] R. Liu *et al.*, "An intriguing failing of convolutional neural networks and the CoordConv solution," *Adv. Neural Inf. Process. Syst.*, vol. 2018-Decem, no. NeurIPS, pp. 9605–9616, 2018.

[21] Y. Zhang, Y. Tian, Y. Kong, B. Zhong, and Y. Fu, "Residual Dense Network for Image Restoration," *IEEE Trans. Pattern Anal. Mach. Intell.*, pp. 1–1, 2020.

[22] S. Woo, J. Park, J. Y. Lee, and I. S. Kweon, "CBAM: Convolutional block attention module," *Lect. Notes*

*Comput. Sci. (including Subser. Lect. Notes Artif. Intell. Lect. Notes Bioinformatics)*, vol. 11211 LNCS, pp. 3–19, 2018.